\documentclass{llncs}

\usepackage{amsmath}

\usepackage{hyperref}
  \hypersetup{
    colorlinks=TRUE,
    citecolor=gray,
    pdftitle ={codeproof},
    linkcolor=gray,
    filecolor=gray,
    urlcolor=gray
    }

\usepackage[color=lightgray]{todonotes}

\usepackage[british]{babel} 
\newcommand{\code}[1]{\texttt{#1}}

\begin{document}

\title{\code{code::proof}: Prepare for \emph{most} weather conditions}
%
%
\author{Charles T. Gray\orcidID{0000-0002-9978-011X}\thanks{
  Thank you to Ben Marwick, Hien Nguyen, Emily Kothe, James Goldie, Mathew Ling, J.D. Long, Kate Smith-Miles, Greg Wilson, Kerrie Mengersen, Jacinta Holloway, Alex Hayes, Noam Ross, Rowland Mosbergen, Luke Prendergast, Dale Maschette, Elio Campitelli, Thomas Lumley, and Daniel S. Katz for advising on particular aspects of this manuscript.
 }}
%
%
\institute{La Trobe University, Melbourne \email{charlestigray@gmail.com}}

\maketitle 
\begin{abstract}

Computational tools for data analysis are being released daily on repositories such as the Comprehensive R Archive Network\footnote{As in the companion manuscript~\cite{gray_truth_2019}, we focus on R packages, but the reader is invited to consider these as examples rather than definitive guidance. The same arguments hold for other languages, such as Python, and associated tools.}. How we integrate these tools to solve a problem in research is increasingly complex and requiring frequent updates. In this manuscript we propose a \emph{toolchain walkthrough}, an opinionated documentation of a scientific workflow. As a practical complement to our proof-based argument~(Gray and Marwick, arXiv, 2019) for reproducible data analysis, here we focus on the practicality of setting up a research compendia with unit tests as a measure of \code{code::proof}, a reproducible research compendia that provides a measure of confidence in computational algorithms.  

\end{abstract}

\section{The Kafkaesque dystopia of DevOps}

In Franz Kafka's 1925 novel \emph{The Trial}~\cite{kafka_trial_2005}, the fictional character Josef K. is prosecuted for crimes that are not clear, in proceedings brought forth by an unidentified authority. For the diligent scientist attempting to answer a mathematical question computationally, such as measuring the efficacy of a statistical estimator via simulation, the process of implementing a scientific workflow to achieve this aim can be a \emph{Kafkaesque} tour of computational tools and systems. The scientist may feel as if they are locked in a dystopia, tested repeatedly for practices in which they have not been trained, such as shell scripts and computational architecture. Whilst there are detailed guides for specific computational tools, it is hard to tell what is still relevant, as code frequently slides into obsolescence~\cite{ragkhitwetsagul_toxic_2019}, and identify the optimal place to begin~\cite{wilson_good_2017}. Significant cultural barriers continue to exist in programming fora; for example,  only one in seventeen contributors to \href{https://stackoverflow.com/}{\underline{Stack Overflow}\footnote{\href{https://stackoverflow.com/}{\underline{Stack Overflow}}~(https://stackoverflow.com/) is forum for asking tightly scoped programming questions.} identify as women~\cite{ford_paradise_2016}}.

For many an unfortunate scientist, the dystopian experience is not confined to the \emph{DevOps}, the developmental operations of preparation for the implementation of an algorithm~\cite{huttermann_devops_2012}. Just as Josef K. was tried multiple times, the labours of the scientist attempting to answer a mathematical question computationally have only just begun. Analogous to how a string will knot with mathematical predictability when jostled~\cite{belmonte_tangled_2007}, an algorithm  will reliably require \emph{debugging}, the process of identifying and correcting code, either to incorporate a new feature, or to correct an error.
This scientist finds themselves part of the first generation of \emph{research software engineers}~(RSEs), who use computational tools in discipline-specific research practices~\cite{wyatt_research_2019}. By virtue of pioneering, RSEs are inadvertently cast as metaresearchers\footnote{
 Visit the \href{https://github.com/softloud/codeproof/issues/2}{\underline{discussion on metaresearch and RSEs}} on the research compendium associated with this manuscript as an example of why this paper, and its companion~\cite{gray_truth_2019}, have so many acknowledgements. Canonical literature is not yet established in the field of RSE, and thus leaders of RSE projects, such as Alex Hayes' maintenance of the \texttt{broom::}~\cite{robinson_broom_2019}. This has propelled Hayes rapidly to the level of expert, by virtue of the pioneering collaborative structure of the package, where hundreds of statistical modellers contribute integrated code.}, developing new methodologies for scientific technologies that hitherto did not exist~\cite{katz_super_2019}. With the aim of mitigating the dystopia of DevOps and debugging for RSEs, this manuscript proposes a \emph{toolchain walkthrough}, an \emph{opinionated}~\cite{parker_opinionated_2017}  documentation of a scientific workflow, towards a measure of \code{code::proof}, a  \emph{good enough}~\cite{wilson_good_2017} effort to provide computational confidence through reproducible research compendia with unit tests.

\section{Toolchain walkthrough}

We define a \emph{toolchain} as a collection of computational tools and commands that forms a scientific workflow to achieve a specific research objective, such as test the efficacy of a statistical estimator in a particular context. The term \emph{walkthrough}, we borrow from video game terminology~\cite{consalvo_zelda_2003}, and is defined as a guide for other players of the game. Various walkthrough formats exist to optimise the narrative enjoyment of the gamer. For example, the Universal Hint System~\cite{uhs_universal_2019} interface provides the gamer with ever more revealing hints without spoiling other parts of the game. Next generation walkthroughs see in-game modifiers, in games such as World of Warcraft, where these provide an option for on-screen boss-specific warnings~\cite{skyisup_deadly_2019}. 

We define \emph{toolchain walkthrough} as an \emph{opinionated}~\cite{parker_opinionated_2017} documentation of a scientific workflow, where opinionated is a term appropriated from software engineering that acknowledges that software guides the user to certain choices. In this manuscript, we describe a workflow for building a research compendium that is opinionated in privileging reproducibility. As with the hint systems of gaming, a workflow can and must be tailored to the skill and background of the user. Thus toolchain walkthroughs can be extended and adapted for different disciplines.

Toolchain walkthroughs have not only intrinsic value in terms of solving the intended research problem, but also extrinsic value, pedagogically and from a developmental perspective. Frequently those who are undertaking research software engineering on statistical projects are not the most senior member of the team; in the case of university faculty, these are often also lecturers and service teachers. There is value in seeing the minutiae of what the footsoldiers of research development undertake and how they instruct others. This can inform as to what skillsets are required in graduate courses, or are required for those who wish to optimise scientific workflow for researchers. Much of what is being implemented right now, in workflows recommended in texts such as \emph{R Packages}~\cite{wickham_r_2015} and \emph{Advanced R}~\cite{wickham_advanced_2014}, is being adopted from existing software engineering principles. Toolchain walkthroughs can contribute to  the literature on the adoption of these procedures in a research context, in addition to programming fora and blog posts.

Blog posts and programming fora, as well as printed texts, are inevitably bound for obsolescence~\cite{ragkhitwetsagul_toxic_2019}. Vignettes, tool-specific long-form documentation~\cite{wickham_r_2015}, focus on one tool in the chain. As a counterpoint to the inadvertently implied redundancy of the academic manuscript in the theoretical companion manuscript~\cite{gray_truth_2019}, here we consider if the ephemerality of most-recent publications, and the chronological nature of academic publishing, may serve the breakneck speed of research development. The toolchain walkthrough provides a documentation of a specific scientific workflow constructed by an expert, or expert in training, in the field. Indeed an expert in training is perhaps best placed, as by virtue of inexperience must research in order to solve the problem. The challenge above, say, the standard one might expect from a blog post, is to provide a \emph{good enough}\footnote{
 As opposed to ofttimes unattainable or impractical \emph{best practices}~\cite{wilson_best_2014} in scientific computing.
}~\cite{wilson_good_2017} effort to avoid questionable research practices~\cite{fraser_questionable_2018} that privilege, say, convention over optimal scientific methodology.

\section{Two research compendia case studies}

For concrete examples of the benefits of adopting software research engineering principals in mathematical science, we consider two in-development research compendia, \code{varameta::} and \code{simeta::}. The primary purpose of these packages is to provide a comparative analysis of estimators for the variance of the sample median when quartiles are provided, rather than a measure of standard deviation, within the meta-analytic context. However, by structuring the packages as such, rather than within a single script file, there is scope for solving similar problems.

\subsection{The \code{varameta::} package; a comparative analysis}

In contemporary meta-analytic computational tools, such as the R package \code{metafor::}~\cite{viechtbauer_conducting_2010}, a measure of both an effect and its variance are required to estimate the population parameters of interest.

However, not all studies report a variance of effect; particularly when scientists suspect an underlying asymmetry in the distribution of the observed data, prompting them to report quartiles, rather than sample standard deviations. One solution to this is to approximate estimators for mean and variance from quartiles~\cite{bland_estimating_2014,hozo_estimating_2005,wan_estimating_2014}. We wish to explore the comparative efficacy of an estimator for the variance of the sample median derived from the estimator of \cite{serfling_approximation_2009}:
$$
 \text{var}(m) \approx \frac{1}{4nf(\nu)^2}
$$
where $m$ denotes the sample median, $n$ the sample size, $\nu$ the population median, and $f$ the population probability density function.

However, in an experimental setting, we do not know the true distribution, nor the true population median. Thu, our method proposes that we assume a distribution, and estimate the parameters of that characterized the assumed distribution from the sample size and sample quartiles. We provide estimators derived for different distributions, to assess the efficacy of this analysis framework. One of which is the exponential distribution, which this manuscript will focus on.

If we assume that $f$ is an exponential probability density function, with unknown rate parameter $\lambda$, then we can estimate this rate parameter via the sample median. Since the true median is given by $\log2/\lambda$, we can estimate the rate parameter,
\begin{equation}
  \label{eqn:exp}
  \lambda \approx \log2 / m,
\end{equation}
via the sample median, $m$.

Each proposed estimator requires a different set of reported values as inputs and different calculations. It is notable that a most optimal estimation method for the problem above is generally unknown. For example, in the comparative analysis  Wan et al.~\cite{wan_estimating_2014}, it was shown that the performance of different estimators varied with the simulated sample sizes.

Thus, there is merit to providing not only the practical functionality of our proposed solution, but also the existing solutions. By structuring this comparative analysis as a reproducible research compendium we achieve practical improvements on a self-contained computational script file. Via \code{roxygen::}ised~\cite{wickham_roxygen2:_2019} documentation, estimators are provided in a modular fashion, with a devoted script file for each estimator that is easily sourced from the package environment. In addition to the advantage of debugging a single script file, the comparative analysis also serves a practical purpose, providing a characterisation of the functionality of each estimator.

To compare these estimators for the variance of the sample median, we undertook \emph{coverage probability} simulations. Here, the coverage probability refers to the probability that the true parameter of interest falls within its constructed confidence interval. In order to do so, we require simulated meta-analytic data, which has the added complexity of a random effect that governs the variation \emph{between} studies. To solve this with confidence in the implementation of computational algorithms and mathematical derivations, we structure this as a package. In addition to building \code{code::proof}, by separating the simulation component, we begin to develop a computational solution to not only solving this problem, but the testing of \code{any} estimator for the variance of the sample mean or median.

\subsection{The \code{simeta::} package}\label{simeta}

A \emph{coverage probability} simulation repeats several  trials with the same simulation meta-parameters where the differing factor is the random sampling of data. In order to separate simulation meta-parameters from trial-level parameters, and delineate this algorithm, we begin by considering a single trial from a standard coverage probability simulation.

\subsection{Coverage probability simulation}

Each trial draws a random sample, for example \code{rnorm(n = 100, mean = 3, sd = 0.2)} will produce 100 values drawn randomly from a normal distribution with mean 3 and standard deviation 0.2. From this sample, we calculate summary statistics. Using these summary statistics, we can compute an estimate of the parameter of interest $\hat\nu$, and its variance $\hat\gamma$. With these estimates, we can produce a $(1-\alpha)\times100\%$ confidence interval $\hat{\nu}\pm z_{1-\alpha/2}\sqrt{\hat{\gamma}}$, where $z_{a}=\Phi^{-1}\left(a\right)$
is the $a\text{th}$ quantile of the standard normal distribution,
and $\Phi$ is the standard normal distribution function. Given we set the parameters for the random sample drawn, we know the true parameter, $\nu$. Thus we can ask, does $\nu$ fall within the confidence interval produced? We summarise the steps of a trial as an algorithm:

\begin{enumerate}
 \item Draw a random sample from the distribution that is characterised by the parameter of interest, $\nu$;
 \item Calculate summary statistics from the random sample;
 \item Calculate an estimate of $\nu$ from the summary statistics;
 \item Construct a confidence interval using the parameter estimate;
 \item Check if $\nu$ falls within the confidence interval.
\end{enumerate}

A coverage probability simulation performs multiple trials and returns the proportion $p \in [0, 1]$ of confidence intervals for $\nu$ that contain the generative parameter value.

\subsection{Simulating meta-analysis data}
\label{sec:sim}

For a meta-analysis simulation, however, these steps are significantly more involved. And with this complexity, as we shall see, nesting, of the algorithm, the advantages of the package structure begin to become apparent. In a single script file, it is hard to find at which step of the algorithm that the code has failed. In addition to human error introduced into code, there are also practical considerations. For example, the random effects maximum likelihood model, \code{method = REML}, employed by \code{metafor::rma}~\cite{viechtbauer_conducting_2010} does not always converge on estimates for the effect and its variance, in which case a fixed effects model, \code{method = FE}, can be employed to produce parameter estimates.

The other point of complexity is in the sampling of meta-analytic data. As meta-analytic data is a collection of summary statistics for $K$ studies of control and intervention samples,  the first step of a coverage probability simulation trial,
\begin{enumerate}
 \item Draw a random sample from the distribution that is characterised by the parameter of interest, $\nu$,
\end{enumerate}
requires several substeps. For the $k\text{th}$ ($k\in\{1,\dots,K\}$) study, we assume there is variation $\gamma_k$ associated with that study, and, in particular, the control, with parameter $\nu_k^C$, and intervention, with parameter $\nu_k^I$, samples with ratio, $\rho = \nu_k^C/\nu_k^I$.

Let us consider a practical example from the estimators provided in the comparative analysis, \code{varameta::}. Our estimator of interest is the variance of the log-ratio of sample medians for control, $\nu^C$, and intervention, $\nu^I$ groups. Since our focus is on building the research compendium to undertake this analysis, rather than the estimators in question, we will take the simplest case, where there is one parameter $\lambda$ associated with the distribution of interest. Let us assume an underlying  exponential distribution: $\text{Exponential}(\lambda)$.

At the simulation level, which is to say, across all trials, we set $\lambda$, the parameter of the distribution of interest. Also at the simulation level, we define a ratio $\rho := \nu^C/\nu^I$ of interest for the population medians, where $\rho = 1$ would indicate no true difference between control and intervention groups. We assume that the log-ratio of sample medians $\log(m^I_k/m^C_k)$ for the $k$th study, can be characterised in terms of the log-ratio of populations medians $\log(\nu^C/\nu^I)$, with some error $\gamma \sim N(0, \tau^2)$ association with that study, as well as sampling error, $\varepsilon \sim N(0, \sigma^2)$,

$$
 \log(m^I_k/m^C_k) = \log(\nu^I/\nu^C) + \gamma_k + \varepsilon_k.
$$

Since the underlying distribution is exponential, we need to find $\lambda_k^J$ for $J \in \{C, I\}$ in order to sample $n$ values $x_1, \dots, x_n \sim  \text{Exponential}(\lambda_k^J)$. We also know the median of the exponential distribution with rate parameter $\lambda$ is given by $\log2/\lambda$. Then, assuming the sampling error will be attained through the random computational process, we have

\begin{align*}
& \log(m_k^I/m_k^C) = \log(\nu^I/\nu^C) + \gamma_k\\
\implies & \log(\lambda_k^C) - \log(\lambda_k^I) = \log(\lambda^C) - \log(\lambda^I) + \gamma_k\\
\implies & \log(\lambda_k^C) - \log(\lambda_k^I) = (\log(\lambda^C) + \gamma_k/2) - (\log(\lambda_k^I) - \gamma_k/2)
\end{align*}

If we then split the random effect associated with the variation between studies $\gamma_k$ equally, and divide the terms by experimental group $J \in \{C, I\}$, we obtain the following system for the control $C$ and intervention $I$ groups' $k$th parameter, $\lambda_k^J$.

\begin{align*}
 \lambda_k^C & = \lambda^C  \exp(\gamma_k/2)\\
 \lambda_k^I & = \lambda^I  \exp(-\gamma_k/2)
\end{align*}

\begin{enumerate}
 \item Draw a measure of variation for the $k$th study from $N(0,\tau^2)$ and calculate $\lambda_I$ from fixed values, the ratio of medians, $\rho$, and the control group's rate parameter $\lambda_C$;
 \item Calculate the rate parameters for the control, $\lambda_k^C$, and intervention, $\lambda_k^I$, groups for the $k$th study;
 \item Draw a random samples of size $n_k^J$ from $\text{Exponential}(\lambda_k^J)$, for $J \in \{C, I\}$.
\end{enumerate}

The sample size $n_k^J$ for the $J$th group of the $k$th study can also be sampled, by assuming $N_k := n_k^C + n_k^I$ and drawing $N_k$ from a uniform distribution $\text{Uniform}(a, b)$, where the minimum $a$, and maximum $b$, reflect knowledge about the domain of interest. The proportion of $N_k$ given to $n_k^I$ can be drawn from a beta distribution. But we shall omit the derivations of these sampling distributions, in the interests of brevity.

In the sampling steps that have been outlined, there are random values drawn, but there are also set simulation-level parameters. We may wish to see how our estimator performs for different numbers of studies, $K$, different expected variability between the studies, $\tau^2$,  and whether or not there is a difference between the control and intervention groups, $\rho$.

And finally, if we consider other distributions, with a mix of symmetric, say, normal or Cauchy distribution, and asymmetric, say, exponential or log-normal, we require different derivations for the sampling parameters.

\subsection{Complexity and formalised analysis structures}

Via the modular nature of a research compendium R package, we can separate each layer of the algorithm into functions. We can produce automated \emph{unit tests} for these functions that, at the very least, check that each component of the algorithm returns an output of expected type. We cannot automate the mathematical derivations, but we can produce an algorithm structure that provides far more computational confidence in implementation than a single script file in which the entire algorithm is nested.

However, structuring an analyses in research compendia is more challenging than simply coding directly into a \code{.R} script. Thus, there is benefit to outlining the computational workflow. We now turn to the practical \emph{toolchain walkthrough} for establishing these analyses as research compendia. We may not be able to prepare for all errors, but we can aim to weather \emph{most} problems that arise in the computational implementation of mathematical algorithms.

\section{Research compendia toolchain walkthrough}

We now aim to provide a practical guide to computational research compendia for the comparative analysis, \code{varameta::}, and the simulation algorithm, \code{simeta::}, that supports it. As this is a first effort at a toolchain walkthrough, there will likely be aspects that are overlooked or underdeveloped.

\subsection{DevOps}

The DevOps section of this toolchain walkthrough aims to cover computational tools, why they were chosen, as well as some guidance as to how to source them.

\subsubsection{Intended audience.}

A toolchain walkthrough is a documentation of a specific scientific workflow created by a scientist who utilised this workflow for research. We begin by identifying the audience targeted who may benefit from detailing the minutiae of this process. We do not seek to generalise, but rather to provide a workflow that reflects the author's knowledge of good enough practices in scientific computing for this task, optimised for efficiency, scientific rigour, and, in the spirit of the gaming walkthrough: \emph{fun}.

This toolchain walkthrough assumes an R user whose expertise is not primarily in computing, but rather a researcher who employs R for analysis in a discipline such as statistics, psychology, archaeology, or ecology. We make an effort to cover some of the less familiar aspects of computational workflow, such as shell commands, that might be considered trivial to a formally trained computer scientist.

Although many R users have gaps in their formal computational science education, researchers who utilise R are often implementing complex algorithms, such as the one outlined in Section \ref{simeta}, which describes the simulation of meta-analysis data for coverage probability simulation.

\subsubsection{Burn it down.} 

This section only applies for work that has already begun. However, this is often the case for the development of a scientific project. We frequently have work that begin as small scripts, that develop in complexity and requirements.

In recognition of the ofttimes overwhelming density of resources, we list a few bash shell commands here that are particularly useful for moving files around when setting up an analysis as a research compendium. We enclose user input in \code{<>} and describe the utility of the command after \code{\#}. A directory is colloquially referred to as a folder. These can be executed from a terminal.


\begin{verbatim}
    . # here
    .. # up one
    cd <directory path> # change location of .
    ls -a # list files in .
    cp <file> <toplace> # copy
    mv <file> <toplace> # move or rename
    rm -rf <directory> # remove directory and its contents
    locate <partoffilename> # find a file
    mkdir <directory> # create a directory
\end{verbatim}

\subsubsection{How to code.}

The R software environment can be downloaded from \href{https://www.r-project.org/}{R: The R Project for Statistical Computing}. There are several excellent resources for getting started with programming with R. We list an opinionated selection here, chosen for clarity and enjoyment, all of which are freely available online:

\begin{itemize}
  \item \href{https://learningstatisticswithr.com/}{Learning Statistics with R} by Danielle Navarro~\cite{navarro_learning_2019},
 \item \href{https://r4ds.had.co.nz/}{R for Data Science} by Grolemund Garrett and Hadley Wickham~\cite{grolemund_r_2017},
 \item \href{https://rc2e.com/}{R Cookbook} by J.D. Long and Paul Teetor~\cite{long_r_2019}.
\end{itemize}

We now assume a working knowledge of the R programming language, as the intended audience of this toolchain workflow are researchers who have a working level of programming proficiency in R.

\subsubsection{Where to code.}

In this toolchain walkthrough, we emphasise cross-platform open-source software. There is, of course, the immediate benefit of accessibility. Furthermore, open-source invites an evolutionary development community where many can contribute small solutions that integrate to solve larger problems. \href{https://www.rstudio.com/}{RStudio} is an integrated development environment for writing in the statistical language \href{https://www.r-project.org/}{R}. RStudio is \emph{cross-platform} in that it can be installed on Windows, Macintosh, and Linux operating systems. There are many further advantages to this widely-used environment. For example, the \code{citr::} add-in~\cite{aust_citr:_2018} modifies RStudio to enable a connection to the open-source reference manager \href{https://www.zotero.org/}{Zotero}. Another example is the \code{datapasta::}~\cite{mcbain_datapasta:_2018} add-in that enables copy-paste of tables into R-formatted script. 

\subsection{Create compendium architecture}

As \code{varameta::} is a research compendium containing comparative analyses and \code{simeta::} a package to provide simulation tools, the creation process for these two compendia are different.

We make use of two R packages,  \code{rrtools::}~\cite{marwick_rrtools_2018} and \code{usethis::}~\cite{wickham_usethis:_2019}, to assist in automating these tasks.

\subsubsection{Compendiumise \code{varameta::}.}

\begin{enumerate}
 \item Open RStudio and close project via the toolbar File menu,
 \item In the Console, set the working directory to desired location; e.g.,
       \begin{verbatim}
        > getwd()
        [1] "/home/charles"
        > setwd("Documents/repos/")
        > getwd()
        [1] "/home/charles/Documents/repos",
    \end{verbatim}
 \item and \code{rrtools::use\_compendium("varameta")},
 \item and update \texttt{DESCRIPTION} file with author, title, etc.,
 \item Create analysis file structure with \code{rrtools::use\_analysis()}.
\end{enumerate}

For \code{varameta::}, we will have several reproducible documents that will form the basis of the analysis, as well as figures to contribute to the associated publication. The final step above automates the creation of a directory structure for a paper, figures, data, and templates.

\subsubsection{Compendiumise \texttt{simeta::}.}

In this case, the file structure is less involved, however the testing structure will be need to be considerably more robust because of the complexity of the simulation algorithm described in Section~\ref{simeta}:

\begin{enumerate}
 \item Create a package with \code{usethis::create\_package()},
 \item Switch to the package directory with \code{usethis::project\_activate()}.
\end{enumerate}

\subsection{Common steps across both packages}

\begin{enumerate}
 \item Set open source licence, with
 \begin{center}
       \code{usethis::use\_mit\_license(name = "Charles Gray")};     
 \end{center}
       this `simple and permissive' choice of licence~\cite{wickham_usethis:_2019} serves the purpose of a comparative analysis of estimators,
 \item Set up documentation for functions with \code{usethis::use\_roxygen\_md()},
 \item Set up data for internal datasets and examples with \code{usethis::use\_data()}.
\end{enumerate}

\subsubsection{Connecting to GitHub.}

There are benefits to implementing a version control system, such as via the Git language and GitHub online repository archive, beyond the ability to trace work back to an earlier iteration~\cite{Bryan2017ExcuseMD}. The added benefit, arguably even greater benefit, is that of collaborative science. Storing work on GitHub allows for instantaneous sharing of code and analyses, and collaborative work with advanced project planning features, enabling other scientists to make very specific comments on work in progress.

\subsubsection{Data ethics and further considerations.}

In the case of \code{varameta::}'s estimators for meta-analysing medians, and \code{simeta::} for simulating meta-analysis estimators, there are no ethics in data considerations beyond ensuring contributors are recognised and credited for their work by time of publication. For some disciplines, sharing geographic locations might be an ethical consideration, say, in preventing fossil hunters from exploiting palaeontology sites~\cite{hopkin_palaeontology_2007}. Personal details, must, too be considered, that might inadvertently identify people and violate privacy considerations. Furthermore, various allowances might need to be made for institutional workflow. We note these here as a possible considerations, but as our case studies do not have such requirements, we now consider our research compendia instantiated. 

However, as this algorithm has significant complexity, we need to include unit tests to provide confidence in our results , as we argue in the companion computational metamathematics manuscript~\cite{gray_truth_2019}, which motivates the practical steps laid out here.  

\section{Testing}

We now expand in a practical sense on unit testing, which, in the theoretical companion manuscript, we describe `the software engineering tool that provides a key piece of the correspondence between scientific claim and programming'~\cite{gray_truth_2019}. It is in this manuscript that we sought to answer the question: why test? In this toolchain walkthrough, we will focus on the practical implementation of first unit tests.

\subsection{What is a test?}

Tests are collected in contexts. Each test comprises congruous expectation functions. 

In the \emph{head} of the `bug hunt' context (under \code{context("bug hunt")}), we find the loading of packages. A seed is then set for reproducibility of errors. The first test, \code{"metasim runs for different n"}, tests the \code{simeta::metasim()} function for different orders of magnitude of \code{trials}. As each trial samples new data, this is the most direct way to test the scalability of the function for large datasets. We then follow up with a test that checks that the exponential distribution can be passed to all levels in the algorithm.

\begin{verbatim}
context("bug hunt")

set.seed(38)
library(tidyverse)
library(metasim)

test_that("metasim runs for different n", {
  expect_is(metasim(), 'data.frame')
  expect_is(metasim(trials = 100) , "data.frame")
  # expect_is(metasim(trials = 1000) , "data.frame")
})

test_that("exponential is parsed throughout", {

  # check sample
  expect_equal(
  sim_sample(10, rdist = "exp", 
    par = list(rate = 3)) %>% length, 10)
  # check samples
  ...
\end{verbatim}

\subsection{Non-empty thing of expected type}

Simply asking `\emph{does a function produce the expected output?}', induces a surprising number of considerations. To illustrate this, we return to our case studies.

\subsubsection{Testing a collection of estimators in \code{varameta::}.}

 In the interests of mathematical and computational brevity, we focus on one distributional example: the simple case of the exponential distribution, which is characterised by a single parameter. We return to the estimator of the rate $\hat\lambda := \log 2 / m$ derived for the exponential distribution, as discussed in Section \ref{sec:sim} and defined in Equation (\ref{eqn:exp}), explicitly coded in R.

\begin{verbatim}
function(n, median) {

  # Estimate parameters.
  lambda <- log(2) / median

  # Approximate the standard error of the sample median.
  1 / (2 * sqrt(n) * dexp(median, rate = lambda))

}
\end{verbatim}

We create a context file, \textbf{tests/testthat/test-exponential.R} and provide a short context description in the first line of the script.

\begin{verbatim}
  context("exponential estimator")
\end{verbatim}

As a starting point, we can write unit tests to automate a check that this function returns non-empty thing of expected type. We arbitrarily choose values, a sample size of 10, and a proposed sample median of 4, for instance. The function should return a numeric \verb|double| value, and should be positive.

\begin{verbatim}
  test_that("non-empty thing of expected type, for fixed values", {

    # returns numeric
    expect_type(g_exp(10, 4), "double")

    # returns positive number
    expect_gt(g_exp(10, 4), 0)

  })
\end{verbatim}

In addition to choosing explicit values, we can also randomly sample the sample size \verb|n|, and sample median \verb|m|. To ensure reproducibility of these testing results on any machine, we set a random seed, passing \verb|set.seed| an arbitrary numeric value.

\begin{verbatim}
set.seed(39) # ensures reproducibility of test results

# sample fuzz testing parameters
n <- sample(seq(2, 100), 1)
m <- runif(1, 1, 100)
\end{verbatim}

We can then use these random \emph{fuzz} values~\cite{klees_evaluating_2018} to produce analogous unit tests for non-empty thing of expected type.

\begin{verbatim}
test_that("non-empty thing of expected type, for random values", {
  expect_type(g_exp(n, m), "double")
  expect_gt(g_exp(n, m), 0)
})
\end{verbatim}

We can extend these tests to cover expected input errors. For example, we wish this function to fail when passed negative numbers. The sample size cannot be less than or equal to 0, and due to the logarithm, the function only works for positive sample medians. Here, we include the fixed and randomised values in the same test.

\begin{verbatim}
  test_that("negative numbers throw an error", {
  expect_error(g_exp(-3, 4))
  expect_error(g_exp(3, -4))
  # with fuzz testing
  expect_error(g_exp(-n, m))
  expect_error(g_exp(n, -m))
})
\end{verbatim}

Running all tests in a context tells us if the function is behaving as expected. The more tests we write, the more confidence we will have that our function behaves as we intended it to.

\begin{verbatim}
==> Testing R file using 'testthat'

Loading varameta
✔ |  OK F W S | Context
✔ |   8       | exponential estimator

══ Results ═════════════════════════════════════════════════
OK:       8
Failed:   0
Warnings: 0
Skipped:  0

Test complete
\end{verbatim}

There is a tradeoff with tests, in terms of time taken by updating the tests themselves. Here a test requires updating from an expected output of a numeric vector, to a dataframe. The function that is being tested. 

\begin{verbatim}
==> Testing R file using 'testthat'

Loading simeta
✔ |  OK F W S | Context
✖ |   6 1     | bug hunt [7.1 s]
──────────────────────────────────────────────────────────────────────────────────────────────────────
test-bug-hunt.R:21: failure: exponential is parsed throughout
sim_stats(rdist = "exp", par = list(rate = 3)) inherits from
`tbl_df/tbl/data.frame` not `numeric`.
──────────────────────────────────────────────────────────────────────────────────────────────────────

══ Results ═══════════════════════════════════════════════════════════════════════════════════════════
Duration: 7.1 s

OK:       6
Failed:   1
Warnings: 0
Skipped:  0

Test complete
\end{verbatim}

\subsubsection{Testing a nested algorithm in \code{simeta::}.}

Our other case study provides an example of a nested algorithm. In addition to ensuring each function returns a non-empty thing of expected type, we can automate checks that the functions form a toolchain. In the first place, it is helpful to know that our functions continue to form a toolchain under default settings.

We begin by setting our context. In this case, as we are running our functions on default settings, we do not require randomly sampled fuzz parmeters.

\begin{verbatim}
  context("default pipeline")
\end{verbatim}

We now check that the algorithm runs `upwards', by running a test from most granular function in the algorithm to most nested. We could write a similiarly inverted test, from most nested function, downwards to most granular.

\begin{verbatim}
  test_that("work upwards through algorithm", {
    expect_is(sim_n(), "data.frame")
    expect_gt(sim_n() %>% nrow(), 1)
    # sim_df calls sim_n
    expect_is(sim_df(), "data.frame")
    expect_is(sim_stats(), "data.frame")
    # metasim calls metatrial
    expect_is(metatrial(), "data.frame")
    expect_is(singletrial(), "data.frame") # alternate trial
    expect_is(metasim(trials = 3), "data.frame")
    # metasims calls sim_df & metasim
    expect_is(metasims(
      single_study = FALSE,
      trials = 3,
      progress = FALSE
    ),
    "sim_ma")
  })
\end{verbatim}

Now, if this test fails, we will know the combination of functions fails at some point in the nested algorithm. We follow this upwards test with a series of small tests for each function set to defaults to identify at which point in the pipeline where the algorithm fails, if the `work upwards' test fails.

\begin{verbatim}
  # test each component on defaults

  test_that("sim_n", {
    expect_is(sim_n(), "data.frame")
  })

  test_that("sim_df", {
    expect_is(sim_df(), "data.frame")
  })

  test_that("metatrial", {
    # metasim calls metatrial
    expect_is(metatrial(), "data.frame")
  })

  test_that("singletrial", {
    expect_is(singletrial(), "data.frame") # alternate trial
  })

  test_that("metasim", {
    expect_is(metasim(trials =  3), "data.frame")
  })

  test_that("metasims", {
    expect_is(metasims(
      single_study = FALSE,
      trials = 3,
      progress = FALSE
    ),
    "list")
  })

\end{verbatim}

And we can now run all tests, for a starting point of automating checks that our algorithm runs on default settings.

\begin{verbatim}
==> Testing R file using 'testthat'

Loading simeta
✔ |  OK F W S | Context
✔ |  14       | default pipeline [28.7 s]

══ Results ═════════════════════════════════════════════════
Duration: 28.7 s

OK:       14
Failed:   0
Warnings: 0
Skipped:  0

Test complete
\end{verbatim}

To demonstrate how informative testing can be in identifying where an algorithm breaks, we now modify the \code{simeta::metasim} function to return a character string, \code{"error"}. Testing the default pipeline reveals where the algorithm is broken. Debugging is where the advantage of testing is exposed, and thus, arguably the requirement for testing increases with complexity of algorithm. Detailed output have been omitted for brevity.

\begin{verbatim}
==> Testing R file using 'testthat'

Loading simeta
✔ |  OK F W S | Context
✖ |  10 4     | default pipeline [32.3 s]
──────────────────────────────────────────────────────────────────────────────────────────────────────
test-default-pipeline.R:12: failure: work upwards through algorithm
metasim(trials = 3) inherits from `character` not `data.frame`.

test-default-pipeline.R:14: error: work upwards through algorithm
Argument 1 must have names
...

test-default-pipeline.R:43: failure: metasim
metasim(trials = 3) inherits from `character` not `data.frame`.

test-default-pipeline.R:47: error: metasims
Argument 1 must have names
...
──────────────────────────────────────────────────────────────────────────────────────────────────────

══ Results ═══════════════════════════════════════════════════════════════════════════════════════════
Duration: 32.3 s

OK:       10
Failed:   4
Warnings: 0
Skipped:  0

Test complete
\end{verbatim}

From this output, we can see not only where the algorithm fails, but also what other functions fail because of a reliance on the elements that have failed.

\subsection{Test-driven development}

As we build new features into our package, such as checking that the single-trial setting works in the simulation function from \code{simeta::}, we can focus on a writing new tests that ensure our feature works within the ecosystem of our algorithm as expected. We can develop our algorithm from a testing setting, rather than focusing on rewriting functions and script files.

Another overview check that we can incorporate is from the \code{covr::} package~\cite{hester_covr_2018}. Using \code{covr::package\_coverage()}, we can check what proportion of lines of code have been tested in each function.

For the \code{varmeta::} package, at the time of writing, we have the following test coverage.

\begin{verbatim}
varameta Coverage: 90.00%
R/g_cauchy.R: 44.44%
R/g_norm.R: 71.43%
R/hozo_se.R: 92.31%
R/bland_mean.R: 100.00%
R/bland_se.R: 100.00%
R/effect_se.R: 100.00%
R/g_exp.R: 100.00%
R/g_lnorm.R: 100.00%
R/hozo_mean.R: 100.00%
R/wan_mean_C1.R: 100.00%
R/wan_mean_C2.R: 100.00%
R/wan_mean_C3.R: 100.00%
R/wan_se_C1.R: 100.00%
R/wan_se_C2.R: 100.00%
R/wan_se_C3.R: 100.00%
\end{verbatim}

This is enables us to target specific functions that may require further testing. Testing lines of code is somewhat a blunt instrument, as we are not ensuring tests for every combination of inputs. However, test coverage is still an informative measure of software reliability. For example, here we see not all code in the \code{g\_*} estimators have been checked.

These notes on testing are not intended to be comprehensive, but only aim to give the user an starting point for the initialisation of summarising an analysis in a reproducible research compendia, with an informative level of automated checks. Given only one quarter of packages on the largest R package repository CRAN have unit tests at all~\cite{gray_truth_2019}, it is arguable that there is much further scope for discussion and development with respect to the adoption of automated tests in reproducible research compendia.

\section{Prepare for \emph{most} weather conditions}

Computational proof may be unachievable, however, a measure of \code{code::proof} can be attained by structuring research compendia in a standardised reproducible format, such as produced by \verb|rrtools::|~\cite{marwick_rrtools_2018}. Perhaps we cannot prove our software in the traditional mathematical sense~\cite{gray_truth_2019}. However, we could consider building confidence in the mathematics that we implement computationally, like waterproofing our shoes. If we step in a big enough puddle, our feet are still going to get wet, but at least we have prepared to weather \emph{most} of the problems associated with the implementation of statistical algorithms.


\bibliographystyle{splncs04}
\bibliography{references.bib}

\end{document}